\documentclass[11pt]{article}

\newcommand{\sect}[1]{\setcounter{equation}{0}\section{#1}}
\newcommand{\subsect}[1]{\subsection{#1}}

\def\be{\begin{equation}}
\def\ee{\end{equation}}
\def\bea{\begin{eqnarray}}
\def\eea{\end{eqnarray}}

\def\bt{\begin{table}}
\def\et{\end{table}}
\def\bc{\begin{center}}
\def\ec{\end{center}}

\def\bbib{\begin{thebibliography}}
\def\ebib{\end{thebibliography}}

\def\1{\'{\i}}

\def\k{\omega}
\def\>#1{{\bf #1}}                 

  \def\z{z}

\begin{document}

\hfill\ 
\bigskip

\begin{center}
{\Large{\bf{On 3+1 anti-de Sitter and de Sitter Lie bialgebras}}}
 
{\Large{\bf{with   
dimensionful deformation parameters}}}
 
\bigskip

{\Large{\bf{}}}

\end{center}

\bigskip

\begin{center}    Angel Ballesteros$^a$, N.~Rossano Bruno$^{a,b}$ and
Francisco~J.~Herranz$^a$\footnote{Communication presented in the XIII Int. Colloq. Integrable 
 Systems and Quantum Groups, June 17-19, 2004, Prague, Czech Republic.}
\end{center}

\begin{center} {\it { 
${}^a$Departamento de F\1sica, Universidad de Burgos, Pza.\
Misael Ba\~nuelos s.n., \\ 09001 Burgos, Spain }}\\ e-mail:
angelb@ubu.es, fjherranz@ubu.es
\end{center}

\begin{center} {\it { 
${}^b$Dipartimento di Fisica, Universit\`a di  Roma Tre  and
INFN Sez.\ Roma Tre,\\ Via Vasca Navale 84, 00146 Roma, Italy}}\\
e-mail: rossano@fis.uniroma3.it
\end{center}

\bigskip\bigskip

\begin{abstract} 
\noindent
We analyze among  all possible   quantum   deformations
of the 3+1 (anti)de Sitter algebras, $so(3,2)$ and $so(4,1)$, which
have  two   specific   non-deformed or primitive commuting operators: the time
translation/energy generator and a     rotation. We prove that
under these conditions there are only two families of two-parametric    
(anti)de Sitter Lie bialgebras. All the deformation parameters appearing in
the bialgebras  are dimensionful ones and they may be related to
   the Planck length. Some properties conveyed by the
corresponding quantum deformations (zero-curvature and 
non-relativistic limits,   space isotropy,\dots) are   studied and their dual
(first-order) non-commutative spacetimes are also presented. 
 \end{abstract}

\newpage


\sect{Introduction}

  Deformed Poincar\'e  symmetries   has recently led  
  to the so called doubly special relativity  
theories~\cite{Amelino-Camelia:2000mn,
MagueijoSmolin,Kowalski-Glikman:2002we,Lukierski:2002df}  which make use
of two fundamental scales:  the usual observer-independent velocity scale (speed of
light $c$) as well as an observer-independent length scale (Planck length $l_p$); the
latter is considered to be related to the deformation parameter. 
Such approaches have established  a direct relationship between quantum groups and
quantum gravity~\cite{amel,KowalskiFS}.    Nevertheless, if Lorentz symmetry has to be
modified at the Planck scale and a non-zero curvature/cosmological constant seems to be
physically relevant, it is necessary  to study the quantum deformations of
the (anti)de Sitter ((A)dS) algebras  covering both the quantum algebra and
its dual quantum group. 

In general, mathematical and physical properties of quantum deformations  
are consequences of the choice of the primitive (non-deformed) generators    in the
quantum algebra. If we require that  one deformation parameter
  plays the  role of a fundamental energy/Planck length (as it seems to be
the case in 3+1 quantum gravity models), then, at least,  the
 energy/time translation generator   $P_0$ must remain primitive. In this way the
uncertainty/dispersion relations could further be    fitted through
corrections depending on $P_0$.  Thus in this paper we study {\em some} of the
possible (first-order) (A)dS deformations (or Lie bialgebras) characterized by a
non-deformed  $P_0$ and analyze their properties.


\sect{(Anti)de Sitter Lie bialgebras}

Let us consider the 3+1 (A)dS algebras algebras   spanned by the generators
of time and space translations    $P_0$ and $P_i$, boosts $K_i$ and rotations $J_i$
$(i=1,2,3)$. Their   commutation rules are collectively denoted by $so_\k(3,2)$ and read
\be
\begin{array}{lll}
[J_i,J_j]=\varepsilon_{ijk}J_k ,& \quad
[J_i,P_j]=\varepsilon_{ijk}P_k  ,&
\quad [J_i,K_j]=\varepsilon_{ijk}K_k ,\\[2pt]
\displaystyle{ [P_i,P_j]=-\frac{\k}{c^2}\,\varepsilon_{ijk}J_k}
,&\quad
\displaystyle{[P_i,K_j]=-\frac{1}{c^2}\,
\delta_{ij}P_0} ,    &\quad
\displaystyle{[K_i,K_j]=-\frac{1}{c^2}\,
\varepsilon_{ijk} J_k} ,\\[6pt]
[P_0,P_i]=\k  K_i ,&\quad [P_0,K_i]=-P_i  ,&\quad
[P_0,J_i]=0 , 
\end{array}
\label{aa}
\ee
where $\k$ is the curvature (proportional to the cosmological constant) of the underlying
classical (A)dS spacetime. If $R$ is the (A)dS radius, the Lie brackets
(\ref{aa}) reduce to  the anti-de Sitter algebra $so(3,2)$ when    $\k=+1/R^2$, and
to the  de Sitter one $so(4,1)$ when    $\k=-1/R^2$. The limit $\k\to 0$ $(R\to \infty)$
in $so_\k(3,2)$ leads to the Poincar\'e algebra, while $c\to \infty$ corresponds to the
non-relativistic contraction.

The most general  classical $r$-matrix, which underlies any
possible quantum deformation of $so_\k(3,2)$,  depends on
45 deformation parameters   $r^{ij}$:
\be
r=r^{ij}Y_i\wedge Y_j=r^{ij}(Y_i\otimes Y_j-Y_j\otimes Y_i) ,
\label{ca}
\ee
where $Y_i$ are the generators of $so_\k(3,2)$. Quantum (A)dS
algebras   are determined, at the
first-order in all  the deformation parameters  $r^{ij}$, by the cocommutator
$\delta$ coming from the $r$-matrix through 
\be
\delta(Y_i)=[Y_i\otimes 1 + 1\otimes
Y_i,r]=f_i^{jk}Y_j\wedge Y_k.
\label{ab}
\ee
  If we require that   $P_0$
remains primitive    $\delta(P_0)=0$ (i.e.\ with full
coproduct $\Delta(P_0)=P_0\otimes 1 + 1\otimes P_0$),  then  the initial 45
deformation parameters are reduced to 15 ones.  Even in this case, the
problem of determining all the
 deformed (A)dS symmetries is a cumbersome   task and  other
considerations must be added.  Thus
we shall impose that another generator which does commute  with $P_0$   remains 
non-deformed as well.   Hence     one of the three rotation  generators will   be taken
primitive, say
$J_3$. The second constraint 
$\delta(J_3)=0$ leaves a  five-parametric candidate $r$-matrix
\bea
&&r=\z_1(K_1\wedge P_1+K_2\wedge P_2)+\z_2(P_1\wedge P_2+\k K_1\wedge K_2)\cr
&&\qquad +\z_3 P_0\wedge
J_3 + \z_4 K_3\wedge P_3+ \z_5 J_1\wedge J_2 ,
\label{cb}
\eea
where $\z_i$ ($i=1,\dots,5$) are our initial set of   deformation parameters which can
also be expressed   as   $\z_i=1/\kappa_i=\ln q_i$.

Next we impose that    the $r$-matrix (\ref{cb}) fulfils the  modified  classical
Yang--Baxter equation for the ten generators $Y_i$ of $so_\k(3,2)$.
In this way  we find five equations:
  \bea
&&  \z_1\z_2=0 ,\nonumber\\ 
&&{c^2}(\z_1-\z_4)\z_5- {\k} \z_2\z_4=0 ,\nonumber\\ 
&&{c^2}\z_2\z_5+  \z_1(\z_1-\z_4)+\k\z_2^2   =0 ,\label{cc}\\[2pt]
&& {c^2}\z_2\z_5-  \z_4(\z_1-\z_4)=0   ,\nonumber\\ 
&&{c^2} \z_5^2+ {\k}\left(\z_2\z_5-\frac{1}{c^2}\,\z_1\z_4    \right)=0,
 \nonumber 
\eea 
which   lead  to   {\em two}  families of two-parametric (A)dS Lie
bialgebras characterized by the following classical $r$-matrices:
\bea
&&\!\!\!\!\!\!\!\!\!\!\!\! r_{\z_1,\z_3}=\z_1\left(K_1\wedge P_1+K_2\wedge
P_2+K_3\wedge P_3 \pm  \frac{\sqrt{\k}}{c^2}\,  J_1\wedge J_2\right) +\z_3 P_0\wedge
J_3    ,\label{cd}\\
 &&\!\!\!\!\!\!\!\!\!\!\!\! r_{\z_2,\z_3}= \z_2\left(P_1\wedge P_2+\k K_1\wedge
K_2-\frac{\k}{c^2}\, J_1\wedge J_2\pm
\sqrt{\k}  P_3\wedge K_3\right) 
 +\z_3 P_0\wedge
J_3   . \label{ce}
\eea
In fact, within each family there are two possibilities according to the sign ``$\pm$". 
The Schouten bracket   of   $r_{\z_l,\z_3}$ $(l=1,2)$   
reads
\bea
&&\!\!\!\!\!\!\!\!\!\!\!\!\!\!\!\!\!\!
 [[ r_{\z_l,\z_3} , r_{\z_l,\z_3} ]]=A_l \left(
\frac{\k}{c^2}\,J_1\wedge J_2\wedge J_3-\frac 12
\varepsilon_{ijk}
\left( \k J_i\wedge K_j\wedge K_k +J_i\wedge P_j\wedge P_k \right ) \right)
\cr
&&\qquad \qquad  +A_l \sum_{i=1}^3
K_i\wedge P_i\wedge P_0 ,\qquad {\rm where}\quad A_1=\frac{\z^2_1}{c^2} ,\quad
A_2=\frac{\z^2_2\k}{c^2} ,
\label{cce}
\eea
which shows that whenever  $\z_1\ne 0$ and $\z_2\ne 0$ both $r$-matrices
would give rise to standard or quasitriangular quantum (A)dS
algebras. On the contrary,    if  
  $\z_1= 0$ and $\z_2= 0$, then   the two $r$-matrices reduce to its common term 
$r_{\z_3}=\z_3 P_0\wedge J_3 $, which is a solution of the classical Yang--Baxter
equation, $[[r_{\z_3},r_{\z_3}]]=0$,  providing a triangular or non-standard (A)dS
deformation   supported by a Reshetikhin twist.

Summing up, each two-parametric  
$r$-matrix and its associated cocommutator has a ``hybrid" character~\cite{BHP} and
can be decomposed as a sum of one standard and another non-standard term: 
\be
r_{\z_l,\z_3}=r_{\z_l}+r_{\z_3} ,\qquad
\delta_{\z_l,\z_3}=\delta_{\z_l}+\delta_{\z_3} ,\qquad l=1,2.
\label{cf}
\ee

  Each one-parametric    (A)dS
bialgebra obtained through
(\ref{ab}) is displayed in Table 1, where hereafter capital Latin indices run as
$I,J=1,2$. Recall that  for all of them
$\delta(P_0)= \delta(J_3)=0$ and that   the complete two-parametric bialgebra,
$(so_\k(3,2),\delta_{\z_l,\z_3}(r))$,  associated to either (\ref{cd}) or
(\ref{ce}) comes from the sum (\ref{cf}). The first one-parametric
bialgebra, $(so_\k(3,2),\delta_{\z_1}(r))$, is of Drinfel'd--Jimbo type and was
obtained in~\cite{LBC}.

We would like to point out that the construction of  the
two-parametric quantum (A)dS algebras (in all orders in $\z_l$ and
$\z_3$) is an open problem, although
 some results concerning the one-parametric deformation with $\z_1$ can be found
in~\cite{karpacz}. In this respect, notice that once a one-parametric quantum
algebra  is obtained for the ``pure" standard component, 
$U_{\z_l}(so_\k(3,2))$,  with coproduct
$\Delta_{\z_l}$ and deformed commutation rules $[Y_i,Y_j]_{\z_l}$, the complete
two-parametric deformation would be provided by the following  twisting element as:
\be
 \Delta_{\z_l,\z_3} = {\cal F}_{\z_3}\Delta_{\z_l} {\cal F}^{-1}_{\z_3} ,\quad
   {\cal F}_{\z_3}=\exp\{-\z_3 P_0\otimes J_3 \} ,\quad
[Y_i,Y_j]_{\z_l,\z_3}\equiv [Y_i,Y_j]_{\z_l}.
\label{cg}
\ee

  
%
%
\bt[t] 
\vspace{-2mm}                         
\caption{(Anti)de Sitter Lie bialgebras $(so_\k(3,2),\delta_{\z_l,\z_3}(r))$
($i,j,k=1,2,3$;
$I,J=1,2$).}
\small
\bc                             
$$
\begin{array}{|l|} 
\hline
 \\[-7pt]
\multicolumn1{|l|}{\quad  \mbox{Standard:}\quad r_{\z_1}=\z_1\left(K_1\wedge
P_1+K_2\wedge P_2+K_3\wedge P_3 \pm  \frac{\sqrt{\k}}{c^2}\,  J_1\wedge J_2\right) } 
\\[6pt]
\displaystyle{\quad \delta_{\z_1}(P_I)=\frac{\z_1}{c^2}\left( P_I\wedge P_0-\k
\varepsilon_{Ijk}J_j\wedge K_k\pm \sqrt{\k}\, J_I\wedge P_3
\right) \quad}  \\[8pt]
\displaystyle{\quad \delta_{\z_1}(P_3)=\frac{\z_1}{c^2}\left( P_3\wedge P_0-\k
\varepsilon_{3jk}J_j\wedge K_k\mp \sqrt{\k}\,(J_1\wedge P_1+J_2\wedge P_2)
\right)  \quad } \\[8pt]
\displaystyle{\quad \delta_{\z_1}(K_I)=\frac{\z_1}{c^2}\left( K_I\wedge P_0+ 
\varepsilon_{Ijk}J_j\wedge P_k\pm \sqrt{\k}\, J_I\wedge K_3
\right) \quad}  \\[8pt]  
\displaystyle{\quad \delta_{\z_1}(K_3)=\frac{\z_1}{c^2}\left( K_3\wedge P_0+ 
\varepsilon_{3jk}J_j\wedge P_k\mp \sqrt{\k}\,(J_1\wedge K_1+J_2\wedge K_2)
\right) \quad}  \\[8pt]
\displaystyle{\quad \delta_{\z_1}(J_I)=\pm \frac{\z_1}{c^2}\, \sqrt{\k}\,  J_I\wedge
J_3 \quad} 
 \\[8pt]
\hline
 \\[-7pt]
\multicolumn1{|l|}{\quad  \mbox{Standard:}\quad  r_{\z_2}= \z_2\left(P_1\wedge P_2+\k
K_1\wedge K_2-\frac{\k}{c^2}\, J_1\wedge J_2\pm
\sqrt{\k}  P_3\wedge K_3\right)  }  \\[6pt]
 \displaystyle{\quad \delta_{\z_2}(P_I)=\frac{\z_2\,\k}{c^2}\left( 
J_3\wedge P_I-J_I\wedge P_3+\varepsilon_{IJ3}K_J\wedge P_0\pm 
\sqrt{\k}\,\varepsilon_{IJ3} J_J\wedge K_3 \right) \quad}  \\[8pt] 
\displaystyle{\quad \delta_{\z_2}(K_I)=\frac{\z_2 }{c^2}\left( 
\k J_3\wedge K_I-\k J_I\wedge K_3-\varepsilon_{IJ3}P_J\wedge P_0\mp 
\sqrt{\k}\,\varepsilon_{IJ3} J_J\wedge P_3 \right) \quad}  \\[8pt] 
\displaystyle{\quad \delta_{\z_2}(P_3)=\pm \frac{\z_2\sqrt{\k}}{c^2}\,
P_0\wedge P_3 \qquad  \delta_{\z_2}(K_3)=\pm \frac{\z_2\sqrt{\k}}{c^2}\,
P_0\wedge K_3 \quad}  \\[8pt] 
\displaystyle{\quad \delta_{\z_2}(J_I)= \z_2 \left( \frac{\k}{c^2}\,
J_3\wedge J_I+P_I\wedge P_3+\k K_I\wedge K_3\pm \sqrt{\k}\,
\varepsilon_{IJ3}\left( K_J\wedge P_3 - P_J\wedge K_3 \right)\right)\quad}   
\\[8pt]
\hline
 \\[-7pt]
\multicolumn1{|l|}{\quad  \mbox{Non-standard:}\quad r_{\z_3}=\z_3 P_0\wedge J_3 } 
\\[4pt]
 \displaystyle{\quad \delta_{\z_3}(P_i)=  \z_3  \left(\k J_3\wedge K_i+
\varepsilon_{ij3} P_j\wedge P_0  \right) \quad}     \\[2pt]
\displaystyle{\quad \delta_{\z_3}(K_i)=-\z_3\left( J_3\wedge P_i-
\varepsilon_{ij3} K_j\wedge P_0  \right) \quad}    \\[2pt]
  \displaystyle{\quad \delta_{\z_3}(J_I)= \z_3 \varepsilon_{IJ3} J_J\wedge P_0  }  
\\[8pt]
\hline
\end{array}
$$
\vspace{-5mm}
\ec                            
\et  

\sect{Properties}

We now study some of the properties which can directly be  derived from the  two
families of (A)dS bialgebras.


\subsect{Dimensions of the deformation  parameters}

 Dimensional analysis of the expressions given in Table 1 shows that the three
parameters $\z_i$ have the following dimensions which are inherited from $P_0$, 
\be
[\z_1]=c^2[P_0]^{-1},\quad [\z_2]=c^2[P_0]^{-2},\quad [\z_3]=[P_0]^{-1}.
\ee
Thus they may play the role of  fundamental scales. In fact, these could
be related to the Planck length $l_{P}$, $z_1\simeq l_{P}$, 
$z_2\simeq l_{\rm P}^2$ and $z_3\simeq l_{\rm P}$, provided
that we consider units with $c=\hbar=1$. Therefore,
these results enable the possibility of working with two possible quantum (A)dS 
deformations, each of them with {\em two}  observer-independent scales (besides $c$).
Obviously within each familiy  both parameters $\z_l$   and $\z_3$ can be
identified (or set proportional) and so reduced to a single fundamental scale.


\subsect{Contractions}

The zero-curvature contraction which starting from  the (A)dS
bialgebras leads to the
  Poincar\'e ones can  be performed by simply applying
the limit  $\k\to \infty$ in the commutation relations (\ref{aa}) and in Table 1. The
resulting expressions are written in Table 2. We remark that
$(iso(3,1),\delta_{\z_1}(r))$ is exactly the well known kappa-Poincar\'e
bialgebra (see~\cite{LukNR} and references therein) so that
  $(iso(3,1),\delta_{\z_1,\z_3}(r))$  is a generalization of   
kappa-Poincar\'e with two deformation parameters. 
The second family $(iso(3,1),\delta_{\z_2,\z_3}(r))$ is now of non-standard type (twisted)
since the Schouten bracket (\ref{cce}) vanishes.
Notice also that a further
non-relativistic limit 
 $c\to \infty$ applied in Table 2 would give  rise
to non-standard Galilean bialgebras.

%
%
\bt[t]                          
\vspace{-2mm}                         
\caption{Contracted Poincar\'e Lie bialgebras  
$(iso(3,1),\delta_{\z_l,\z_3}(r))$.}
\small
\bc                             
$$
\begin{array}{|l|} 
\hline
 \\[-7pt]
\multicolumn1{|l|}{\quad   \mbox{Standard:}\quad  r_{\z_1}=\z_1\left(K_1\wedge
P_1+K_2\wedge P_2+K_3\wedge P_3  \right)   } 
\\[6pt]
\displaystyle{\quad \delta_{\z_1}(P_i)=\frac{\z_1}{c^2}\left( P_i\wedge P_0 
\right) \quad\  \delta_{\z_1}(K_i)=\frac{\z_1}{c^2}\left( K_i\wedge P_0+ 
\varepsilon_{ijk}J_j\wedge P_k 
\right) \quad\  \delta_{\z_1}(J_i)=0\quad} 
 \\[8pt]
\hline
 \\[-7pt]
\multicolumn1{|l|}{\quad  \mbox{Non-standard:}\quad   r_{\z_2}= \z_2 P_1\wedge P_2   } 
\\[4pt]
 \displaystyle{\quad \delta_{\z_2}(P_i)=0 \quad\ \delta_{\z_2}(K_I)=-\frac{\z_2 }{c^2}\,
\varepsilon_{IJ3}P_J\wedge P_0   \quad\    \delta_{\z_2}(K_3)=0 \quad\  \delta_{\z_2}(J_I)=
\z_2    P_I\wedge P_3  \quad}   
\\[8pt]
\hline
 \\[-7pt]
\multicolumn1{|l|}{\quad  \mbox{Non-standard:}\quad  r_{\z_3}=\z_3 P_0\wedge J_3 }  \\[4pt]
 \displaystyle{\quad \delta_{\z_3}(P_i)=  \z_3    
\varepsilon_{ij3} P_j\wedge P_0   \qquad \delta_{\z_3}(J_I)= \z_3 \varepsilon_{IJ3} J_J\wedge
P_0  \quad}    \\[2pt]
  \displaystyle{\quad 
\delta_{\z_3}(K_i)=-\z_3\left( J_3\wedge P_i-
\varepsilon_{ij3} K_j\wedge P_0  \right) 
 }  
\\[8pt]
\hline
\end{array}
$$
\vspace{-5mm}
\ec                              
\et  

\subsect{Non-commutative (anti)de Sitter spacetimes}

If   $\hat y^i$ denotes the    quantum group coordinate dual to
$Y_i$   such that  $\langle \hat y^i| Y_j\rangle=    \delta_j^i$,    
   Lie bialgebra     duality   provides the 
 first-order quantum group  $[\hat y^i,\hat y^j]=f_k^{ij}\hat y^k$ dual to
the bialgebra (\ref{ab}). In our case,  let   $\{ \hat\theta, \hat
x_\mu,  \hat\xi_i \} $ be the  non-commutative coordinates dual to the 
generators $\{J,P_\mu,K_i\}$, respectively. Then the  non-commutative spacetimes
associated to the bialgebras of Table 1 arise as  the
commutation rules involving the quantum coordinates $\hat x_\mu$, namely
\bea
r_{\z_1,\z_3}:&& [\hat x_0,\hat x_i]=-\frac{\z_1}{c^2}\,\hat
x_i  +\z_3
\,\varepsilon_{ij3}\, \hat x_j  ,\qquad [\hat x_i,\hat x_j]=0.
\label{fa}\\
r_{\z_2,\z_3}:&&
 [\hat x_0,\hat
x_I]=-\frac{\z_2}{c^2}\,\varepsilon_{IJ3}\,\hat
\xi_J    +\z_3
\,\varepsilon_{IJ3}\, \hat x_J  , \quad   [\hat x_0,\hat
x_3]=\pm\frac{\z_2}{c^2}\,\sqrt{\k}\,\hat x_3 , \nonumber\\
&&    [\hat
x_I,\hat x_3]= \z_2 \hat\theta_I ,\qquad  [\hat x_1,\hat x_2]=0  .
\eea
We stress that the kappa-Minkowski space~\cite{LukNR} 
appears within   (\ref{fa}) when $\z_3=0$ and that  such a
first-order structure is simultaneously shared  by the (A)dS and Poincar\'e
cases as there is no   dependence on  the curvature $\k$. Nevertheless, further
corrections   depending on $\k$ can be expected to appear in the complete
quantum (A)dS spaces similarly to   what happens in the 2+1 case 
\cite{BBH}.


\subsect{Space isotropy}

The requirement that  one rotation generator (as $J_3$)   remains primitive from the
beginning of our approach   suggests that, in principle,   space isotropy is
intrinsically lost at the level of the quantum algebra as the different
formal expressions given in Table 1 for
$J_3$ and
$\{J_1,J_2\}$ explicitly   show. However we consider that the ``space isotropy
condition" deserves a deeper analysis and that   the most natural object
to impose it is the   non-commutative spacetime   which must be  
covariant under quantum group rotations. Hence this point could   be checked
once the complete quantum algebra and the dual (all-orders) non-commutative
spacetime were obtained. In this respect,  see~\cite{ads} for an explicit
example where a new non-commutative Minkowskian spacetime associated to a
quantum Weyl--Poincar\'e algebra is shown to be covariant under quantum
group transformations.


\bigskip
{\small This work was partially supported  by the Ministerio de
Educaci\'on y Ciencia   (Spain, Project FIS2004-07913), by the
Junta de Castilla y Le\'on   (Spain, Project  BU04/03), and by the INFN-CICyT
(Italy-Spain).}
\bigskip


\end{document}